\documentclass[wssquare]{ws-procs975x65} 
\begin{document}
\title{Spectral asymptotics for $\delta'$ interaction supported by a infinite curve}

\author{Michal Jex}

\address{Doppler Institute for Mathematical Physics and Applied
Mathematics, \\ Czech Technical University in Prague,
B\v{r}ehov\'{a} 7, 11519 Prague, \\ and Department of Physics,
Faculty of Nuclear Sciences and Physical Engineering, Czech
Technical University in Prague, B\v{r}ehov\'{a} 7, 11519 Prague,
Czech republic\\ $^*$E-mail: jexmicha@fjfi.cvut.cz}

\begin{abstract}
We consider a generalized Schr\"odinger operator in $L^2(\mathbb
R^2)$ describing an attractive $\delta'$ interaction in a strong
coupling limit. $\delta'$ interaction is characterized by a coupling
parameter $\beta$ and it is supported by a $C^4$-smooth infinite
asymptotically straight curve $\Gamma$ without self-intersections.
It is shown that in the strong coupling limit, $\beta\to 0_+$, the
eigenvalues for a non-straight curve behave as $-\frac{4}{\beta^2}
+\mu_j+\mathcal O(\beta|\ln\beta|)$, where $\mu_j$ is the $j$-th
eigenvalue of the Schr\"odinger operator on $L^2(\mathbb R)$ with
the potential $-\frac14 \gamma^2$ where $\gamma$ is the signed
curvature of $\Gamma$.
\end{abstract}


\bodymatter

\section{Introduction}
The quantum mechanics describing the particle confined to various
manifolds is studied quite extensively. It is very useful for
describing various nanostructures in physics but it also offers a
large variety of interesting problems from the purely mathematical
point of view. Systems where the confinement is realized by a
singular attractive potential, so called 'leaky' quantum graphs
\cite{Ex}, have the advantage that they take quantum tunneling
effects into account in contrast to quantum graphs \cite{BK}. The
confining potential is often taken to be of the $\delta$ type. One
can think also about more singular types of potentials namely
$\delta'$ type based on the concept of $\delta'$ interaction
in one dimension \cite{AGHH}.\\

We are interested in the spectrum of the operator which can be
formally written as
\begin{equation*}\label{operator}
H=-\Delta-\beta^{-1}\delta'(\cdot-\Gamma)
\end{equation*}
where $\delta'$ interaction is supported by an infinite curve
$\Gamma$ in $\mathbb R^2$. We are interested in the strong coupling
regime which corresponds to small values of the parameter $\beta$.
We derive spectral asymptotics of discrete and essential spectra. As
a byproduct we obtain that for a non-straight curve the bound state
arises for sufficiently small $\beta$ in an alternative way to one
presented in \cite{BEL}.

\section{Formulation of the Problem and Results}
We consider a curve $\Gamma$ parameterized by its arc length
\begin{equation*}
\Gamma:\:\mathbb R\rightarrow\mathbb{R}^2\,, \quad
s\mapsto(\Gamma_1(s),\Gamma_2(s))\,,
\end{equation*}
where $\Gamma_1(s),\Gamma_2(s)\in C^4(\mathbb R)$ are component
functions. We denote signed curvature as $\gamma(s) := (\Gamma''_1
\Gamma'_2 - \Gamma'_1 \Gamma''_2)(s)$. We introduce several conditions for the curve $\Gamma$ as:\\
($\Gamma1$) $\Gamma$ is $C^4$ smooth curve,\\
($\Gamma2$) $\Gamma$ has no ``near self-intersections'', i.e. there
exists its strip neighborhood of a finite thickness which does not
intersect with itself,\\
($\Gamma3$) $\Gamma$ is asymptotically straight in the sense that
$\lim_{|s|\rightarrow\infty}\gamma(s)=0$ and\\
($\Gamma4$) $\Gamma$ is not a straight line.\\

The operator, we are interested in, acts as a free Laplacian outside
of the interaction support
\begin{equation*}
(H_{\beta}\psi)(x)=-(\Delta\psi)(x)
\end{equation*}
for $x\in\mathbb{R}^2\setminus\Gamma$ with the domain which can be
written as $\mathcal{D}(H_\beta)=\{\psi\in
H^2(\mathbb{R}^2\setminus\Gamma) \mid
\partial_{n_\Gamma}\psi(x)=\partial_{-n_\Gamma}\psi(x)=\psi'(x)|_{\Gamma},\,
-\beta\psi'(x)|_{\Gamma}=\psi(x)|_{\partial_+\Gamma}
-\psi(x)|_{\partial_-\Gamma}\}$. The vector $n_\Gamma$ denotes the
normal to $\Gamma$ and $\psi(x)|_{\partial_\pm\Gamma}$ are the
appropriate traces of the function $\psi$. For the purpose of the
proofs we introduce curvelinear coordinates $(s,u)$ along the curve
in the same way as done in \cite{EJ}, i.e.
\begin{equation} \label{curvilin}
(x,y)=\big(\Gamma_1(s)+u\Gamma'_2(s),\Gamma_2(s)-u\Gamma'_1(s)\big)\,.
\end{equation}
As a result of the conditions ($\Gamma1$) and ($\Gamma2$) it can be
shown that the map (\ref{curvilin}) is injective for all $u$ small
enough. We denote $d$ as a maximum for which the map
(\ref{curvilin}) is injective. A strip neighborhood around $\Gamma$
of thickness $a<d$ is denoted by $\Omega_a:= \{x\in\mathbb R^2:\,
\mathrm{dist\,}(x,\Gamma)<a\}$.\\

The quadratic form associated with the operator $H_\beta$ was
derived in \cite{BLL} and it can be written as
\begin{equation*}
h_\beta[\psi]= \|\nabla \psi\|^2 -\beta^{-1}\int_{\mathbb
R}|\psi(s,0_+)-\psi(s,0_-)|^2\,\mathrm d s\,.
\end{equation*}
where we used the curvelinear coordinates in the strip neighborhood
of the curve $\Gamma$ for the functions $\psi\in C(\mathbb R^2) \cap
H^1(\mathbb{R}^2\setminus\Gamma)$ as $\psi(s,u)$. We also need to
introduce the operator defined on the line as
\begin{equation} \label{comparison}
S=-\frac{\partial^2}{\partial s^2}-\frac{1}{4}\gamma(s)^2\,,
\end{equation}
with the domain $\mathcal D(S)=H^2(\mathbb R)$. The eigenvalues of
the operator $S$ are denoted by $\mu_j$ with the multiplicity taken
into account. Now we are ready to write down the main results of our
paper.

\begin{theorem}\label{thm1}
Let an infinite curve $\Gamma$ satisfy conditions
$(\Gamma1)$--$(\Gamma3)$, then $\sigma_\mathrm{ess}(H_\beta)
\subseteq [\epsilon(\beta),\infty)$, where $\epsilon(\beta)\to
-\frac{4}{\beta^{2}}$ holds as $\beta\to 0_+$.
\end{theorem}

\begin{theorem}\label{thm2}
Let an infinite curve $\Gamma$ satisfy assumptions
$(\Gamma1)$--$(\Gamma4)$, then $H_\beta$ has at least one isolated
eigenvalue below the threshold of the essential spectrum for all
sufficiently small $\beta>0$, and the $j$-th eigenvalue behaves in
the strong coupling limit $\beta\to 0_+$ as
\begin{equation*}
\lambda_j=-\frac{4}{\beta^2}+\mu_j+\mathcal O(-\beta\ln(\beta))\,.
\end{equation*}
\end{theorem}

\section{Bracketing estimates}
For the proofs of both theorems we will need estimates of our
operator $H_\beta$ via Dirichlet and Neumann bracketing as done in
\cite{EY}. We introduce the operators with added either Dirichlet or
Neummann boundary conditions at the boundary of the strip
neighborhood $\Omega_a$ of $\Gamma$. We introduce quadratic forms
$h^+_{\beta}$ and $h^-_{\beta}$ on the strip neighborhood of
$\Gamma$ which can be written as
\begin{equation*}
h^\pm_{\beta}[\psi]= \|\nabla \psi\|^2 -\beta^{-1}\int_{\mathbb
R}|\psi(s,0_+)-\psi(s,0_-)|^2\,\mathrm d s
\end{equation*}
with the domains $\mathcal
D(h^+_{\beta})=\tilde{H}^1_0(\Omega_a\setminus\Gamma)$ and $\mathcal
D(h^-_{\beta})=\tilde{H}^1(\Omega_a\setminus\Gamma)$. The operators
associated with the quadratic forms $h^\pm_{\beta}$ are denoted by
$H^\pm_{\beta}$, respectively. With the help of Dirichlet-Neumann
bracketing we are able to write the following inequality
\begin{equation}
\label{nerovnost} -\Delta_{\mathbb R^2\setminus\Omega_a}^N\oplus
H^-_{\beta}\leq H_\beta\leq-\Delta_{\mathbb
R^2\setminus\Omega_a}^D\oplus H^+_{\beta}\,,
\end{equation}
where $-\Delta_{\mathbb R^2\setminus\Omega_a}^{N,D}$ denotes either
Neumann or Dirichlet Laplacian on $\mathbb R^2\setminus\Omega_a$
respectively. Neumann Laplacian and Dirichlet Laplacian are positive
and as a result all the information about the negative spectrum,
which we are
interested in, is encoded in the operators $H^\pm_{\beta}$.\\
Now we rewrite the quadratic forms $h^\pm_{\beta}$ in the
curvelinear coordinates \eqref{curvilin}. We obtain expression which
are analogical to those obtained in \cite{EJ}, i.e.
\begin{lemma}
Quadratic forms $h^+_{\beta}$, $h^-_{\beta}$ are unitarily
equivalent to quadratic forms $q^+_{\beta}$ and $q^-_{\beta}$ which
can be written as
\begin{eqnarray*}
q^+[f]= \|\frac{\partial_s f}{g}\|^2 +\|\partial_u f\|^2 +(f,Vf)
-\beta^{-1}\int_{\mathbb
R}|f(s,0_+)-f(s,0_-)|^2\, \mathrm d s \\
+ \frac12 \int_{\mathbb
R} \gamma(s) \big(|f(s,0_+)|^2-|f(s,0_-)|^2\big)\, \mathrm d s \\[.3em]
q^-[g]= q_D[g] -\int_{\mathbb
R}\frac{\gamma(s)}{2(1+a\gamma(s))}|f(s,a)|^2\, \mathrm d s
+\int_{\mathbb R}\frac{\gamma(s)}{2(1-a\gamma(s))}|f(s,-a)|^2\,
\mathrm d s
\end{eqnarray*}
defined on $\tilde{H}^1_0(\mathbb R \times ((-a,0)\cup (0,a)))$ and
$\tilde{H}^1(\mathbb R \times ((-a,0)\cup (0,a)))$, respectively.
The geometrically induced potential in these formul{\ae} is given by
\begin{equation*}
V(s,u)=\frac{u\gamma''}{2g^3} -\frac{5(u\gamma')^2}{4g^4}
-\frac{\gamma^2}{4g^2}
\end{equation*}
with $g(s,u):= 1+u\gamma(s)$.
\end{lemma}
The proof of this lemma can be done step by step as done in
\cite{EJ} so we omit the details.\\

We will also need cruder estimates by quadratic forms
$b^\pm_\beta[f]$ which satisfy $b^-_\beta[f]\leq q^-[f]\leq
h_\beta[f]\leq q^+[f]\leq b^+_\beta[f]$. The quadratic form
$b^+_\beta[f]$ can be written as
\begin{eqnarray*}
b^+_\beta[f]=\|\partial_u f\|^2 +(1-a\gamma_+)^{-2}\|\partial_s
f\|^2 +(f,V^{(+)}f) \\ -\beta^{-1}\int_{\mathbb
R}|f(s,0_+)-f(s,0_-)|^2 \mathrm d s +\frac12 \int_{\mathbb R}
\gamma(s) \big(|f(s,0_+)|^2-|f(s,0_-)|^2\big)\, \mathrm d s
\end{eqnarray*}
where
$V^{(+)}:=\frac{a(\gamma'')_+}{2(1-a\gamma_+)^3}-\frac{\gamma^2}{4(1+a\gamma_+)^2}$
and $f_+:=\max_{s\in\mathbb R} |f|$ denotes maximum of $|f|$. The
quadratic form $b^-_\beta[f]$ can be written as
\begin{eqnarray*}
b^-_\beta[f]=\|\partial_u f\|^2+(1+a\gamma_+)^{-2} \|\partial_s
f\|^2 +(f,V^{(-)}f) \\ -\beta^{-1}\int_{\mathbb
R}|f(s,0_+)-f(s,0_-)|^2 \,\mathrm d s - \frac12 \int_{\mathbb
R}\gamma(s)
\big(|f(s,0_+)|^2-|f(s,0_-)|^2\big) \,\mathrm d s \\
-\gamma_+\int_{\mathbb R}|f(s,a)|^2\,\mathrm d
s-\gamma_+\int_{\mathbb R}|f(s,-a)|^2\,\mathrm d s
\end{eqnarray*}
where $V^{(-)}=-\frac{a(\gamma'')_+}{2(1-a\gamma_+)^3}
-\frac{5(a(\gamma')_+)^2}{4(1-a\gamma_+)^4}
-\frac{\gamma^2}{4(1-a\gamma_+)^2}$. The operators $B^\pm_\beta$
associated with $b^\pm_\beta[f]$ can be written as
 $B^\pm_\beta=U_a^\pm\otimes I+\int_{\mathbb R}^{\oplus} T_{a,\beta}^{\pm}(s)\,\mathrm d
 s$ where $U_a^\pm$ corresponds to the longitudinal variable $s$ and
 $T_{a,\beta}^{\pm}(s)$ corresponds to the transversal variable $u$.
The operators $T_{a,\beta}^{\pm}(s)$ act as $T_{a,\beta}^{\pm}(s)f
 =-f''$ with the domains
 \begin{eqnarray*}
\hspace{-2em} \lefteqn{\mathcal{D}(T_{a,\beta}^{+}(s))= \big\{f\in
H^2((-a,a)\setminus\{0\})\mid f(a)=f(-a)=0\,,} \\ &&
f'(0_-)=f'(0_+)= -\beta^{-1}(f(0_+)-f(0_-))+\frac12
\gamma(s)(f(0_+)+f(0_-)) \big\}\\
\hspace{-2em} \lefteqn{\mathcal{D}(T_{a,\beta}^{-}(s))= \big\{f\in
H^2((-a,a)\setminus\{0\})\mid\: \mp\gamma_+f(\pm a)=f'(\pm a)\,,} \\
&& f'(0_-)=f'(0_+)= -\beta^{-1}(f(0_+)-f(0_-))+\frac12
\gamma(s)(f(0_+)+f(0_-)) \big\}\,.
\end{eqnarray*}
The operators $U_a^\pm$ act as $U_a^\pm f=-(1\mp
a\gamma_+)^{-2}f''+V^{(\pm)}f$ with the domain
$\mathcal{D}(U_a^\pm)=H^2(\mathbb R)$. The operators
$T_{a,\beta}^{\pm}(s)$ depend on the
 variable $s$, however, their negative spectrum is independent of
 $s$. Now we state two lemmata estimating the eigenvalues of
 operators $T_{a,\beta}^{\pm}(s)$ and $U_a^\pm$. Their proofs can be found in \cite{EJ} so we omit
the details.

\begin{lemma}\label{trans}
Each of the operators $T_{a,\beta}^{\pm}(s)$ has exactly one
negative eigenvalue $t_\pm(a,\beta)$, respectively, which is
independent of $s$ provided that $\frac{a}{\beta}>2$ and
$\frac{2}{\beta}>\gamma_+$. For all $\beta>0$ sufficiently small
these eigenvalues satisfy the inequalities
\begin{equation*}
-\frac{4}{\beta^2}-\frac{16}{\beta^2}\exp\left(-\frac{4a}{\beta}\right)\leq
t_-(d,\beta)\leq-\frac{4}{\beta^2}\leq
t_+(d,\beta)\leq-\frac{4}{\beta^2}+\frac{16}{\beta^2}\exp\left(-\frac{4a}{\beta}\right)\,.
\end{equation*}
\end{lemma}

\begin{lemma} \label{long}
There is a positive $C$ independent of $a$ and $j$ such that
\begin{equation*}
|\mu_j^\pm(a)-\mu_j|\leq Caj^2
\end{equation*}
holds for $j\in\mathbb N$ and $0<a<\frac{1}{2\gamma_+}$, where
$\mu_j^\pm(a)$ are the eigenvalues of $U_a^\pm$, respectively, with
the multiplicity taken into account.
\end{lemma}

Now we are ready to prove our main theorems.

\section{Proof of Theorem \ref{thm1}}
First we prove the trivial case for the straight line. By separation
of variables the spectrum is $\sigma(H_\beta) =
\sigma_\mathrm{ess}(H_\beta)
= \big[-\frac{4}{\beta^2},\infty\big)$.\\

The case for non-straight curve is done similarly as for the
singular interaction supported by nonplanar surfaces in
\cite{EK,EJ2}. The inclusion $\sigma_\mathrm{ess}(H_\beta) \subseteq
[\epsilon(\beta),\infty)$ can be rewritten as
\begin{equation*}
\inf\sigma_\mathrm{ess}(H_\beta) \geq \epsilon(\beta)\,.
\end{equation*}
The inequality $H_\beta\geq H^-_\beta\oplus -\Delta_{\mathbb
R^2\setminus\Omega_d}^N$ implies that it is sufficient to check
$\inf\sigma_\mathrm{ess}(H^-_\beta) \geq \epsilon(\beta)$ in
$L^2(\Omega_a)$ for $a<d$ because the operator $-\Delta_{\mathbb
R^3\setminus\Omega_d}^N$ is positive. Next we divide the curve
$\Gamma$ into two parts. First part is defined as
$\Gamma^{\mathrm{int}}_\tau:=\{\Gamma(s)|s<\tau\}$ and the second
one
$\Gamma^{\mathrm{ext}}_\tau:=\Gamma\setminus\overline{\Gamma^{\mathrm{int}}_\tau}$.
The corresponding strip neighborhoods are defined as
$\Omega_a^{\mathrm{int}}:=\{x(s,u)\in\Omega_a|s<\tau\}$ and
$\Omega_a^{\mathrm{ext}}:=\{x(s,u)\in\Omega_a|s>\tau\}$. We
introduce Neumann decoupled operators on
$\Omega_a^{\mathrm{int},\mathrm{ext}}$ as
\begin{equation*}
H_{\beta,\tau}^{-,\mathrm{int}}\oplus
H_{\beta,\tau}^{-,\mathrm{ext}}\,.
\end{equation*}
The operators $H_{\beta,\tau}^{-,\omega}$,
$\omega=\mathrm{int},\mathrm{ext}$ are associated with quadratic
forms $h_{\beta,\tau}^{-,\omega}$ which can be written as
\begin{eqnarray*}
h_{\beta,\tau}^{-,\omega}= \|\frac{\partial_s f}{g}\|^2
+\|\partial_u f\|^2 +(f,Vf) -\beta^{-1}\int_{\Gamma_\tau^{\omega}}|f(s,0_+)-f(s,0_-)|^2\, \mathrm d s\\
+ \frac12 \int_{\Gamma_\tau^{\omega}} \gamma(s)
\big(|f(s,0_+)|^2-|f(s,0_-)|^2\big)\, \mathrm d s\\
-\int_{\Gamma_\tau^{\omega}}\frac{\gamma(s)}{2(1+a\gamma(s))}|f(s,a)|^2\,
\mathrm d s
+\int_{\Gamma_\tau^{\omega}}\frac{\gamma(s)}{2(1-a\gamma(s))}|f(s,-a)|^2\,
\mathrm d s
\end{eqnarray*}
with the domains $\tilde H^{1}(\Omega_a^{\omega})$. Neumann
bracketing implies that $H_{\beta,\tau}^{-}\geq
H_{\beta,\tau}^{-,\mathrm{int}}\oplus
H_{\beta,\tau}^{-,\mathrm{ext}}$. The spectrum of the operator
$H_{\beta,\tau}^{-,\mathrm{int}}$ is purely discrete \cite{Da} and
as a result min-max principle implies that
\begin{equation*}
\inf\sigma_\mathrm{ess}(H_{\beta,\tau}^{-}) \geq
\inf\sigma_\mathrm{ess}(H_{\beta,\tau}^{-,\mathrm{ext}})\,.
\end{equation*}
We denote the following expression
$V_{\tau}:=\inf_{|s|>\tau,u\in(-a,a)}V(s,u)$. The assumption
$(\Gamma2)$ gives us that
\begin{eqnarray*}
\lim_{\tau\rightarrow\infty}V_{\tau}=0
\end{eqnarray*}
With the help of Lemma \ref{trans} we can write the following
estimates
\begin{eqnarray*}
h_{\beta,\tau}^{-,\mathrm{ext}}[f]\geq
\|\partial_u f\|^2 +V_{\tau}\|f\|^2-\beta^{-1}\int_{\Gamma_\tau^{\mathrm{ext}}}|f(s,0_+)-f(s,0_-)|^2\, \mathrm d s\\
+ \frac12 \int_{\Gamma_\tau^{\mathrm{ext}}} \gamma(s)
\big(|f(s,0_+)|^2-|f(s,0_-)|^2\big)\, \mathrm d s\\
-\int_{\Gamma_\tau^{\mathrm{ext}}}\frac{\gamma(s)}{2(1+a\gamma(s))}|f(s,a)|^2\,
\mathrm d s
+\int_{\Gamma_\tau^{\mathrm{ext}}}\frac{\gamma(s)}{2(1-a\gamma(s))}|f(s,-a)|^2\,
\mathrm d s\\
\geq
\left(V_{\tau}-\frac{4}{\beta^2}-\frac{16}{\beta^2}\exp\left(-\frac{4a}{\beta}\right)\right)\|f\|^2
\end{eqnarray*}
Because we can choose $\tau$ arbitrarily large we obtain the the
desired result.

\section{Proof of Theorem \ref{thm2}}
For the proof of the second theorem we use the inequalities
\eqref{nerovnost} and Lemmata \ref{trans} and \ref{long}. First we
put $a(\beta)=-\frac{3}{4}\beta\ln\beta$. Now with the explicit form
of $B^\pm_\beta$ in mind and the fact that $T_{a,\beta}^{\pm}(s)$
have exactly one negative eigenvalue we have that the spectra of
$B^\pm_\beta$ can be written as
$t_\pm(d(\beta),\beta)+\mu_j^{\pm}(d(\beta))$. Using Lemmata
\ref{trans} and \ref{long} we obtain
\begin{equation*}
t_\pm(a(\beta),\beta)+\mu_j^{\pm}(a(\beta))=-\frac {4}
{\beta^{2}}+\mu_j+\mathcal O(\beta|\ln\beta|)\,.
\end{equation*}
The min-max principle along with the inequality \eqref{nerovnost}
completes the proof.

\subsection*{Acknowledgments}

\noindent The research was supported by the Czech Science Foundation
within the project \mbox{14-06818S} and by by Grant Agency of the
Czech Technical University in Prague, grant No.
SGS13/217/OHK4/3T/14.


\end{document}